\documentstyle[prl,aps,twocolumn,floats,epsf]{revtex}

\newcommand{\postbb}[3]
{\setlength{\epsfxsize}{#3\hsize}
 \centerline{\epsfbox[#1]{#2}}}

\newcommand{\epc}[2]{{\em Eur. Phys. J.}            {\bf C#1}, #2 }

\newcommand{\etal}{{\em et al.}}

\newcommand{\be}{\begin{equation}}
\newcommand{\ee}{\end{equation}}
\newcommand{\ba}{\begin{array}}
\newcommand{\ea}{\end{array}}

\newcommand{\lsim}{\buildrel < \over {_\sim}}

\begin{document}

\preprint{
\noindent
\hfill
\begin{minipage}[t]{3in}
\begin{flushright}
UPR--T--XXX \\
\vspace*{2cm}
\end{flushright}
\end{minipage}
}

\draft

\title{The Probability Density of the Mass of the Standard Model Higgs Boson}
\author{Jens Erler}
\address{Department of Physics and Astronomy, University of Pennsylvania, 
Philadelphia, PA 19104-6396, USA}

\date{October 2000}

\maketitle

\begin{abstract}
The LEP Collaborations have reported a small excess of events in their combined
Higgs boson analysis at center of mass energies $\sqrt{s} \lsim 208$~GeV. 
In this communication, I present the result of a calculation of the probability
distribution function of the Higgs boson mass which can be rigorously 
obtained if the validity of the Standard Model is assumed.  It arises from
the combination of the most recent set of precision electroweak data and 
the current results of the Higgs searches at LEP~2.
\end{abstract}

\pacs{PACS numbers: 14.80.Bn, 12.15.Mm, 12.15.Ji.}

Combining all Higgs decay channels and experiments, the LEP Collaborations 
report a $2.5\sigma$ excess in their data~\cite{Junk00}. The probability that 
this is due to an upward fluctuation of the background is 0.6\%. Of course, 
the Higgs boson has been searched for in many different energy bins, and there 
is an infinitely large energy range out of reach, so that one expects 
to observe an upward fluctuation somewhere. It is therefore difficult 
to interpret these numbers, and it would be imprudent to conclude that 
the Higgs boson has been found with 99.4\% probability.

In this communication, I present the answer to a different but related 
question, which is {\sl Given the data, what is the probability that the Higgs 
boson is within reach of LEP~2?}  This question can be answered unambiguously 
once the probability distribution of the Higgs boson mass, $M_H$, has been 
constructed. This is not possible given the Higgs search results at LEP~2 by 
themselves, regardless of how strong a signal is observed there. The reason is 
that there is an infinite domain of {\em a priori\/} possible values of $M_H$ 
beyond the kinematic reach of LEP~2. As a result, the $M_H$ distribution is 
improper, {\em i.e.,} it is asymptotically non-zero. Including the electroweak
precision data, however, renders it sufficiently convergent and a proper 
integration is possible. 

\begin{figure}[t]
\postbb{20 60 530 690}{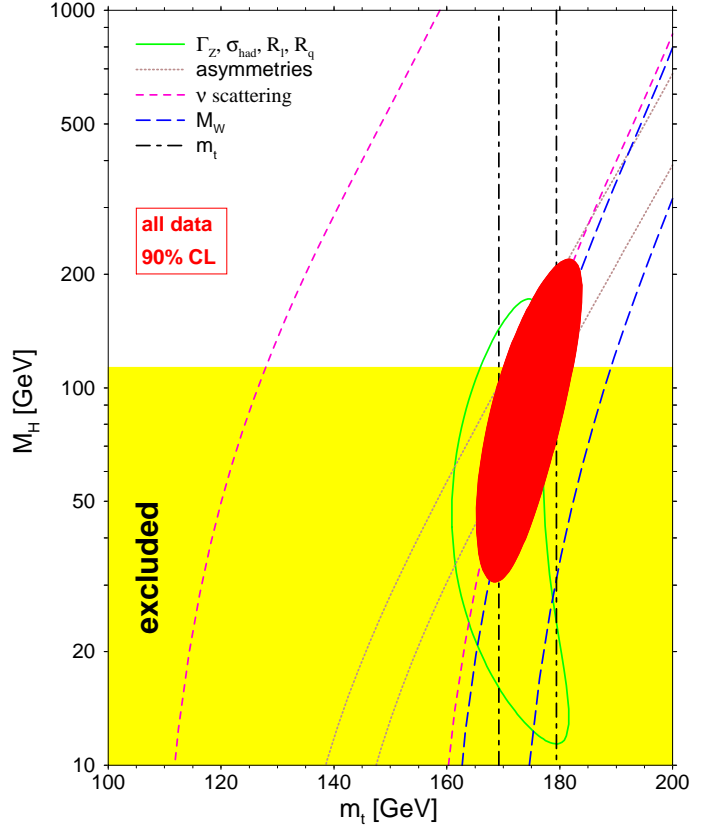}{1.12}
\caption{ Constraints on $M_H$ from various sets of precision data as 
a function of the top quark mass. For the individual data sets I show $1\sigma$
contours, while the ellipse for all data refers to the 90\% CL. The excluded
region by LEP~2, $M_H < 113.2$~GeV (95\% CL), is also indicated.}
\label{fig1}
\end{figure}

The electroweak precision data by itself, especially $Z$ pole asymmetries and 
the $W$ boson mass, is now precise enough to constrain $M_H$ significantly, as
can be seen from Fig.~\ref{fig1}.
Including the latest updates as presented at the 2000 summer conferences, 
I find from a global fit to all data using the package GAPP~\cite{GAPP},
\be
\label{mh}
   M_H = 86^{+48}_{-32} \mbox{ GeV}. 
\ee
Note, that by definition the central value in Eq.~(\ref{mh}) maximizes 
the likelihood, $N e^{-\chi^2 (M_H)/2}$, and that correlations with other 
parameters, $\xi^i$, are accounted for, since minimization w.r.t.\ these 
is understood, $\chi^2 \equiv \chi^2_{\rm min}$. The error is the standard
$1\sigma$ uncertainty ($\Delta \chi^2 = 1$). The 68.27\% {\em central 
confidence interval\/}, 
\be
\label{indirect68}
   53 \mbox{ GeV} \leq M_H \leq 131 \mbox{ GeV},
\ee
differs slightly from Eq.~(\ref{mh}) due to the non-Gaussian (asymmetric)
distribution of $\ln M_H$. The 90\% central confidence interval yields,
\be
\label{indirect}
   38 \mbox{ GeV} \leq M_H \leq 173 \mbox{ GeV}.
\ee
The right hand side of Eq.~(\ref{indirect}), {\it i.e.,} the 95\% upper limit, 
does not take into account the direct search results of the LEP Collaborations.
Negative search results will increase the upper limit, because the probability 
distribution function is effectively being renormalized. For example, in 
a previous analysis~\cite{Erler00} we found that the Higgs exclusion curve 
presented by the LEP Collaborations increased the 95\% upper limit by 30 GeV.

The use of the Higgs exclusion curve, however, is only appropriate if no
indication of an excess is observed.  In general, it is more appropriate
to consider the likelihood ratio for the data,
\be
   Q_{\rm LEP~2} = \frac{{\cal L} ({\rm data|signal + background})}
                        {{\cal L} ({\rm data|background})},
\ee
where both the numerator and the denominator are functions of $M_H$. 
The quantity, 
\be
   \Delta \chi^2 ({\rm direct}) \equiv - 2 \ln Q_{\rm LEP~2}
\ee
can then be added to the $\chi^2$-function obtained from the precision data. 
If the signal hypothesis gives a better (worse) description of the data than 
the background only hypothesis we find a negative (positive) contribution
to the total $\chi^2$. Note, that this is a consistent treatment also in 
the case of a large downward fluctuation of the background or even if no events
are observed at all. Use of $Q_{\rm LEP~2}$ had been originally advocated
in Ref.~\cite{Degrassi99} (see Eq.~(23) in that reference).

This treatment can be rigorously justified within the framework of Bayesian
statistics~\cite{Bayes1763,Gelman95}, which is particularly suited for 
parameter estimation. Bayesian methods are based on Bayes 
theorem~\cite{Bayes1763},
\be
\label{Bayes}
  p(M_H|{\rm data}) = \frac{p({\rm data}|M_H) p(M_H)}{p({\rm data})},
\ee
which must be satisfied once the likelihood, $p({\rm data}|M_H)$, and 
{\em prior\/} distribution, $p(M_H)$, are specified.
$p(data) \equiv \int p({\rm data}|M_H) p(M_H) d M_H$ in the denominator
provides the proper normalization of the {\em posterior\/} distribution on
the left hand side. Depending on the case at hand, the prior can
\begin{enumerate}
 \item contain additional information not included in the likelihood model,
 \item contain likelihood functions obtained from previous measurements,
 \item or be chosen {\em non-informative}.
\end{enumerate}
Of course, the posterior does not depend on how information is separated into
the likelihood and the prior. As for the present case, I choose 
the {\em informative\/} prior,
\be
  p(M_H) = Q_{\rm LEP~2}~p^{\rm non-inf}(M_H),
\ee
where the non-informative part of the prior will be chosen as
\be
  p^{\rm non-inf}(M_H) = M_H^{-1}.
\ee
This choice corresponds to a flat prior in the variable $\ln M_H$, and there
are various ways to justify it~\cite{Gelman95}. One rationale is that a flat
distribution is most natural for a variable defined over all the real numbers. 
This is the case for $\ln M_H$ but not $M_H^2$. Also, it seems that 
{\em a priori\/} it is equally likely that $M_H$ lies, say, between 30 and 
40~GeV, or between 300 and 400~GeV. In any case, the sensitivity of 
the posterior to the (non-informative) prior diminishes rapidly with 
the inclusion of more data. As discussed before, $p(M_H)$ is an improper prior 
but the likelihood constructed from the precision measurements will provide 
a proper posterior.

Occasionally, the Bayesian method is criticized for the need of a prior, which 
would introduce unnecessary subjectivity into the analysis. Indeed, care and 
good judgement is needed, but the same is true for the likelihood model, which 
has to be specified in any statistical model. Moreover, it is appreciated among
Bayesian practitioners, that the explicit presence of the prior can be 
advantageous: it manifests model assumptions and allows for sensitivity checks.
From the theorem~(\ref{Bayes}) it is also clear that any other method must 
correspond, mathematically, to specific choices for the prior. Thus, Bayesian 
methods are more general and differ rather in attitude: by their strong 
emphasis on the entire posterior distribution and by their first principles 
setup. 

Including $Q_{\rm LEP~2}$ in this way, one obtains the 95\% CL upper limit 
$M_H \leq 201$~GeV, {\em i.e.\/} notwithstanding the observed excess events, 
the information provided by the Higgs searches at LEP~2 {\em increase\/} 
the upper limit by 28~GeV.

Given extra parameters, $\xi^i$, the distribution function of $M_H$ is defined 
as the marginal distribution, 
$p(M_H|{\rm data}) = \int p(M_H, \xi^i | {\rm data}) \prod_i p(\xi^i) d \xi^i$.
If the likelihood factorizes, $p(M_H, \xi^i) = p(M_H) p(\xi^i)$, the $\xi^i$ 
dependence can be ignored. If not, but $p(\xi^i | M_H)$ is 
(approximately) multivariate normal, then
$$ \chi^2 (M_H,\xi^i) = \chi^2_{\rm min} (M_H) + $$ \vspace{-20pt}
$$ {1\over 2} \frac{\partial^2 \chi^2 (M_H)} {\partial \xi_i \partial \xi_j} 
   (\xi^i - \xi^i_{\rm min} (M_H)) (\xi^j - \xi^j_{\rm min} (M_H)). $$
The latter applies to our case, where $\xi^i = (m_t,\alpha_s,\alpha(M_Z))$. 
Integration yields, 
\be
  p(M_H | {\rm data}) \sim \sqrt{\det E}\, e^{- \chi^2_{\rm min} (M_H)/2},
\ee
where the $\xi^i$ error matrix, $E = (\frac{\partial^2 \chi^2 (M_H)}
{\partial \xi_i \partial \xi_j})^{-1}$, introduces a correction factor
with a mild $M_H$ dependence. It corresponds to a shift relative to the 
standard likelihood model, 
$\chi^2 (M_H) = \chi^2_{\rm min}(M_H) + \Delta \chi^2 (M_H)$, where
\be
  \Delta \chi^2 (M_H) \equiv \ln \frac{\det E (M_H)}{\det E (M_Z)}.
\ee
This effect {\em tightens} the $M_H$ upper limit by 1~GeV. 

I also include theory uncertainties from uncalculated higher orders. 
This increases the upper limit by 5~GeV, 
\be
   M_H \leq 205 \mbox{ GeV} \mbox{ (95\% CL)}.
\ee

\begin{figure}[t]
\postbb{80 120 530 690}{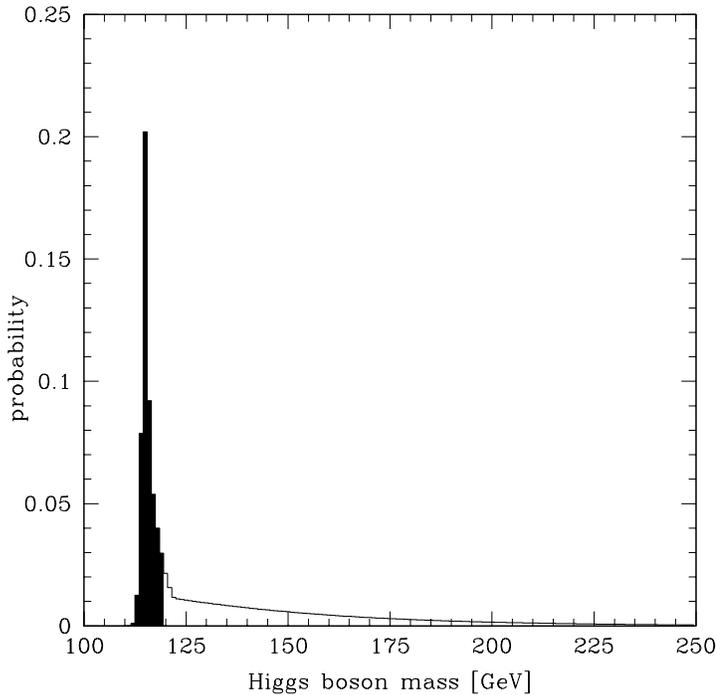}{0.88}
\caption{ Probability distribution function for the Higgs boson mass.
The probability is shown for bin sizes of 1~GeV. Included are all available 
direct and indirect data. The shaded and unshaded regions each mark 50\%
probability.}
\label{fig2}
\end{figure}

The entire probability distribution is shown in Fig.~\ref{fig2}. Taking
the data at face value, there is (as expected) a significant peak around 
$M_H = 115$~GeV, but more than half of the probability is for Higgs boson 
masses above the kinematic reach of LEP~2 (the median is at $M_H = 119$~GeV). 
However, if one would double the integrated luminosity and assume that 
the results would be similar to the present ones, one would find most of 
the probability concentrated around the peak. A similar statement will apply
to Run II of the Tevatron at a time when about 3 to 5 ${\rm fb}^{-1}$ of data 
have been collected.

The described method is robust within the SM, but it should be cautioned that 
$M_H$ extracted from the precision data is strongly correlated with certain 
new physics parameters.  Likewise, the Higgs searches at LEP~2 depend on 
the predictions of signal and background expectations which are strictly
calculable only within a specified theory. This note focussed on the Standard 
Model Higgs boson. 

\centerline{\bf Acknowledgements:}
This work was supported in part by the US Department of Energy grant 
EY--76--02--3071.

\end{document}